\documentclass{arxivpaperproc}
\usepackage{latexsym}
\usepackage{mathptmx}

	\usepackage[english]{babel}
	\usepackage{amsmath}
	\usepackage{amssymb}
 	\usepackage{amsthm}
	\usepackage{bbm}
	\usepackage{xypic}
	\usepackage[all]{xy}
	\usepackage[pdftex]{graphicx}
	\usepackage[abs]{overpic}
	\usepackage{algorithm}
	\usepackage{subfigure}
	\usepackage{upgreek}
	\usepackage{textcomp}
	\usepackage{mathrsfs}
	\usepackage{leftidx}
	\usepackage{bigints}
	\usepackage{abbrevs}
	\usepackage{placeins}
\usepackage[pdftex,colorlinks=true,urlcolor=blue,citecolor=black,anchorcolor=black,linkcolor=black]{hyperref}

    \newcommand{\mvbar}{\middle\vert}
	\newcommand{\ud}{\,\mathrm{d}}
	\newcommand{\tm}{\,\!_{t\text{-}}} 

	\newcommand{\U}{\text{U}}

	\newcommand{\Poi}{\text{Poi}}

    \newcommand{\fin}[1]{{#1}^{\text{fin}}}
    \newcommand{\rem}[1]{{#1}^{\text{rem}}}

    \newcommand{\algref}[1]{\hyperref[#1]{Algorithm \ref*{#1}}}
    \newcommand{\stepref}[1]{\hyperref[#1]{Step \ref*{#1}}}
    \newcommand{\algstref}[2]{\hyperref[#2]{Algorithm \ref*{#1} Step \ref*{#2}}}
    \newcommand{\figref}[1]{\hyperref[#1]{Figure \ref*{#1}}}
    \newcommand{\subfigref}[3]{\hyperref[#1]{Figure \ref*{#2}(#3)}}
    \newcommand{\tabref}[1]{\hyperref[#1]{Table \ref*{#1}}}
    \newcommand{\apxref}[1]{\hyperref[#1]{Appendix \ref*{#1}}}
    \newcommand{\secref}[1]{\hyperref[#1]{Section \ref*{#1}}}
    \newcommand{\prinref}[1]{\hyperref[#1]{Principle \ref*{#1}}}
    \newcommand{\conref}[1]{\hyperref[#1]{Condition \ref*{#1}}}
    \newcommand{\resref}[1]{\hyperref[#1]{Result \ref*{#1}}}
    \newcommand{\defnref}[1]{\hyperref[#1]{Definition \ref*{#1}}}
    \newcommand{\thmref}[1]{\hyperref[#1]{Theorem \ref*{#1}}}
    \newcommand{\lemref}[1]{\hyperref[#1]{Lemma \ref*{#1}}}
    \newcommand{\corref}[1]{\hyperref[#1]{Corollary \ref*{#1}}}
    \newcommand{\remref}[1]{\hyperref[#1]{Remark \ref*{#1}}}

    \newname\rnd{Radon-Nikod{\'y}m derivative}

\newtheoremstyle{wsc}
{3pt}
{3pt}
{}
{}
{\bf}
{}
{.5em}
{}

\theoremstyle{wsc}
\newtheorem{theorem}{Theorem}

\newtheorem{defn}{Definition}
\newtheorem{cond}{Condition}
\newtheorem{prin}{Principle}
\newtheorem{res}{Result}

\begin{document}

\pagestyle{fancyplain}

\thispagestyle{plain}
\firstPageHead{}

\chead{\fancyplain{}{\itshape }}

\rhead{}
\cfoot{}
\renewcommand{\headrulewidth}{0pt} 

\input{wscbib.tex}           

\setlength{\baselineskip}{12.7pt}

\title{On the Exact Simulation of (Jump) Diffusion Bridges}
\author{Murray Pollock\\ [12pt]
Department of Statistics\\
University of Warwick \\
Gibbet Hill Road\\
Coventry, CV4 7AL, UK}

\maketitle

\section*{ABSTRACT}
In this paper we outline methodology to \textit{efficiently} simulate (jump) diffusion bridge sample paths \textit{without discretisation error}. We achieve this by considering the simulation of conditioned (jump) diffusion bridge sample paths in light of recent work developing a mathematical framework for simulating finite dimensional sample path \textit{skeletons} (which flexibly characterise the entirety of sample paths).

\section[INTRODUCTION]{INTRODUCTION}
\label{s:introduction}

Diffusions and jump diffusions are an important class of stochastic processes widely used to model phenomena in a broad range of application areas, such as economics and finance \cite{JPE:BS73} and the life sciences \cite{SC:GW06}. Diffusions are also widely used throughout computational statistics as their simulation underpins a broad class of highly efficient Markov chain Monte Carlo algorithms \cite{B:RT96}. A jump diffusion $V:\,\mathbbm{R}\!\to\!\mathbbm{R}$ is a Markov process which can be defined as the solution to a stochastic differential equation (SDE) of the form (denoting $V_{t-} := \lim_{s \uparrow t }V_s$), 
\begin{align}
\ud V_t 
& = \beta(V\tm) \ud t + \sigma(V\tm) \ud W_t + \ud J^{\lambda,\mu}_t,\quad\quad V_0 = v \in\mathbbm{R},\,\,\,t\in[0,T], \label{eq:jump diffusion}
\end{align}
where $\beta:\,\mathbbm{R}\!\to\!\mathbbm{R}$ and $\sigma:\,\mathbbm{R}\!\to\!\mathbbm{R}_+$ denote the (instantaneous) drift and diffusion coefficients respectively, $W_t$ is a standard Brownian Motion and $J^{\lambda,\mu}_t$ denotes a compound Poisson process. $J^{\lambda,\mu}_t$ is parameterised with (finite) jump intensity $\lambda:\,\mathbbm{R}\!\to\!\mathbbm{R}_+$ and jump size coefficient $\mu:\,\mathbbm{R}\!\to\!\mathbbm{R}$ with jumps distributed with density $f_{\mu}$. All coefficients are themselves (typically) dependent on $V_t$ and regularity conditions are assumed to hold to ensure the existence of a unique non-explosive weak solution (see \cite{BK:ASCJD}).

We may naturally be interested in simulating sample paths from the measure on the path space induced by (\ref{eq:jump diffusion}), which we denote by $\mathbbm{T}^{v}_{0,T}$. Clearly this is non trivial as sample paths are infinite dimensional random variables (and so at most we can simulate some finite dimensional subset of sample paths) and, as a closed form representation of the transition density of (\ref{eq:jump diffusion}) will be typically unavailable, we may need to resort to time \textit{discretisation} \cite{BK:NSSDE} which results in the introduction of error. To address these challenges a class of rejection-sampling based algorithms (so called \textit{Exact Algorithms} as they avoid the introduction of error) have been developed to simulate a broad range of diffusions \cite{AAP:BR05,MCAP:BPR08,MOR:CH13,arxiv:J13,TR:JS14} and jump diffusions \cite{MCAP:CR10,MCAP:GR13,B:PJR15} by means of simulating from an equivalent measure $\mathbbm{P}^{v}_{0,T}$.

In this paper we construct exact algorithms to tackle the related problem of simulating conditioned jump diffusion sample paths, which can be represented as the solution to an SDE of the following form,
\begin{align}
\ud V_t 
& = \beta(V\tm) \ud t + \sigma(V\tm) \ud W_t + \ud J^{\lambda,\mu}_t,\quad\quad V_0 = v \in\mathbbm{R},\,\,\,V_T = w \in\mathbbm{R},\,\,\,t\in[0,T].  \label{eq:conditioned diffusion}
\end{align}
A conditioned jump diffusion (or \textit{jump diffusion bridge}) is simply a diffusion which in addition to having a given start point is also conditioned to have some specified end point. For the purposes of this paper we restrict our attention to univariate diffusions and impose a number of additional conditions on the coefficients of (\ref{eq:jump diffusion},\ref{eq:conditioned diffusion}) (as detailed in \secref{s:preliminaries}).

As in (\ref{eq:jump diffusion}), we are interested in simulating sample paths from the measure induced by (\ref{eq:conditioned diffusion}), denoted $\mathbbm{T}^{v,w}_{0,T}$, which (as outlined in \cite{PhD:P13,MCAP:GR13}) can be achieved by constructing an equivalent measure $\mathbbm{P}^{v,w}_{0,T}$ from which sample paths can be drawn.  There are two key complications when constructing an exact algorithm to simulate conditioned jump diffusions which are not present in simulating unconditioned jump diffusions \cite{PhD:P13}. Firstly, construction of an appropriate equivalent measure $\mathbbm{P}^{v,w}_{0,T}$ is more difficult. Secondly, the computational cost of simulating  conditioned (jump) diffusions does not necessarily scale linearly as a function of the time interval in which it has to be simulated over, and so exact algorithms can be rendered computationally infeasible for particular applications.

In this paper we outline methodology to simulate conditioned jump diffusion sample paths, employing strategies to accelerate acceptance and rejection of proposal sample paths and reduce overall computational cost. We achieve this by considering the simulation of conditioned jump diffusions in light of recent work developing a mathematical framework for simulating diffusion sample path \textit{skeletons} (characterising the entirety of sample paths), and the extension of exact algorithms to \textit{Adaptive Exact Algorithms}  (which enable the simulation of lower dimensional skeletons) \cite{B:PJR15}.

This paper is organised as follows: In \secref{s:preliminaries} we introduce the key concepts, framework and conditions imposed in establishing the results presented in this paper. In \secref{s:cauea} we introduce more formally exact algorithms and introduce a novel adaptive exact algorithm for simulating conditioned diffusions. Finally, in \secref{s:caujea} we extend our approach to simulating conditioned jump diffusions. 

\section[PRELIMINARIES]{PRELIMINARIES} \label{s:preliminaries}
In \cite{B:PJR15} a framework for constructing exact algorithms was established in which entire (jump) diffusion sample paths could be represented by means of simulating a finite dimensional \textit{skeleton}, guided by three key principles. The skeleton typically comprises a \textit{layer} constraining the sample path. In this section we will begin by reviewing these definitions and principles for exact algorithms below, and then outline the notation and conditions imposed to establish the results in this paper.
\begin{defn}[Skeleton] \label{defn:skeleton}
A skeleton $(\mathcal{S})$ is a finite dimensional representation of a diffusion sample path $(V\sim\mathbbm{T}^{v,w}_{0,T})$, that can be simulated without any approximation error by means of a proposal sample path drawn from an equivalent proposal measure $(\mathbbm{P}^{v,w}_{0,T})$ and accepted with probability proportional to $\frac{\ud\mathbbm{T}^{v,w}_{0,T}}{\ud\mathbbm{P}^{v,w}_{0,T}}(V)$, which is sufficient to restore the sample path at any finite collection of time points exactly with finite computation where $V|\mathcal{S}\sim \left.\mathbbm{P}^{v,w}_{0,T}\mvbar\mathcal{S}\right.$.
\end{defn}

\begin{defn}[Layer] \label{defn:layer}
A layer $R(V)$, is a function of a diffusion sample path $V\sim\mathbbm{P}^{v,w}_{0,T}$ which determines the compact interval to which any particular sample path $V(\omega)$ is constrained. 
\end{defn}

\begin{prin}[Layer Construction] \label{prin:layer}
The path space of the process of interest, can be partitioned and the layer to which a proposal sample path belongs can be unbiasedly simulated, $R(V)\sim\mathcal{R} := \mathbbm{P}^{v,w}_{0,T} \circ R^{-1}$.
\end{prin}
\begin{prin}[Proposal Exactness] \label{prin:prop}
Conditional on $V_0$, $V_T$ and $R(V)$, we can simulate any finite collection of intermediate points of the trajectory of the proposal diffusion exactly, $V\sim \left.\mathbbm{P}^{v,w}_{0,T}\mvbar_{R^{-1}\left(R(V)\right)}\right.$.
\end{prin}
\begin{prin}[Path Restoration] \label{prin:rest}
Any finite collection of intermediate (inference) points, conditional on the skeleton, can be simulated exactly, $V_{t_1},\ldots{},V_{t_n}\sim\left.\mathbbm{P}^{v,w}_{0,T}\mvbar\mathcal{S}\right.$.
\end{prin}

To present our work in some generality we assume Conditions \ref{cond:exis}--\ref{cond:cjea} hold. A fuller discussion of the conditions imposed can be found in \cite[\textsection 1.3, \textsection 4.2 \& \textsection 5.4]{PhD:P13}.
\begin{cond}[Solutions] \label{cond:exis}
The coefficients of (\ref{eq:jump diffusion},\ref{eq:conditioned diffusion})  are sufficiently regular to ensure the existence of a unique, non-explosive, weak solution.
\end{cond}
\begin{cond}[Continuity] \label{cond:cont}
The drift coefficient $\beta \in C^1$. The volatility coefficient $\sigma\in C^2$ and is strictly positive.
\end{cond}
\begin{cond}[Growth Bound] \label{cond:grow}
We have that $\exists\,K>0$ such that $|\beta(x)|^2 + |\sigma(x)|^2 \leq K(1+|x|^2)$ $\forall x\in\mathbbm{R}.$
\end{cond}
\begin{cond}[Jump Rate] \label{cond:jump}
$\lambda$ is non-negative and there exists a constant $\Lambda<\infty$ such that $\lambda\leq\Lambda$.
\end{cond}

Conditions \ref{cond:cont} and \ref{cond:grow} are sufficient to allow us to transform our SDEs in (\ref{eq:jump diffusion},\ref{eq:conditioned diffusion}) into one with unit volatility (letting $\psi_1,\ldots{},\psi_{N_T}$ denote the jump times in the interval $[0,T]$, $\psi_0:=0$ and $\psi_{N_T+1}-:=\psi_{N_T+1}:=T$, and $N_t := \sum_{i \geq 1} \mathbbm{1}\{\psi_i \leq t\}$ a Poisson jump counting process). As noted in \cite{AS:A08}, this transformation is typically possible for univariate diffusions and for many multivariate diffusions.
\begin{res}[Lamperti Transform \protect{\cite[Chap. 4.4]{BK:NSSDE}}] \label{res:lamp}
Let $\eta(V_t) =: X_t$ be a transformed process, where $\eta(V_t) := \int^{V_t}_{v^*} 1/\sigma(u) \ud u$ (where $v^*$ is an arbitrary element in the state space of $V$), then by applying It\^o's formula for jump diffusions to find $\ud X_t$ we have (where $\mu_t\sim f_{\mu}(\cdot;V\tm) = f_{\mu}(\cdot;\eta^{-1}(X\tm))$),
\begin{align} 
\ud X_t
& = \underbrace{\left[\dfrac{\beta\left(\eta^{-1}\!\left(X\tm\right)\right)}{\sigma\left(\eta^{-1}\!\left(X\tm\right)\right)}\!-\!\dfrac{\sigma'\left(\eta^{-1}\!\left(X\tm\right)\right)}{2}\right]}_{\alpha\left(X\tm\right)} \ud t\!+\!\ud W_t\!+\!\underbrace{\left(\eta\!\left[\eta^{-1}\!\left(X\tm\right) + \mu_t\right]\!-\!X\tm\right)\ud N_t}_{\ud J^{\lambda,\nu}_t}.\label{eq:dXt}
\end{align}
\end{res}

We denote the measure induced by the transformed unconditioned jump diffusion (\ref{eq:dXt}) as $\mathbbm{Q}^x_{0,T}$ (with left hand point $X_0 := x = \eta(v)$), and $\mathbbm{Q}^{x,y}_{0,T}$ as the measure induced by the transformed conditioned jump diffusion (constrained to have end point $X_T := y = \eta(w)$). We further denote by $\mathbbm{W}^x_{0,T}$ as the measure induced by the following driftless jump diffusion with unit volatility,
\begin{align}
\ud X_t 
& = \ud W_t + \ud J^{\Lambda,\delta}_t, \quad\quad X_0 = x \in\mathbbm{R},\,\,\, t\in[0,T], \label{eq:w radon diff}
\end{align}
where $J^{\Lambda,\delta}_t$ is a compound Poisson process with constant finite jump intensity $\Lambda$, jump size coefficient $\delta:\,\mathbbm{R}\!\to\!\mathbbm{R}$ and with jumps distributed with density $f_{\delta}$. We denote by $J_{[0,T]}$ as the trajectory of a compound Poisson process over $[0,T]$ and $\mathbbm{W}^{x,y}_{0,T}$ as the measure induced by (\ref{eq:w radon diff}) where we additionally have $X_T=y$. 

In order to deploy an exact algorithm we need to establish that the \rnd of $\mathbbm{Q}^{x,y}_{0,T}$ with respect to $\mathbbm{W}^{x,y}_{0,T}$ exists (Results \ref{res:radon} and \ref{res:transition}) and can be bounded on compact sets (\resref{res:compact}). In order to do so we impose on the coefficients of (\ref{eq:dXt},\ref{eq:w radon diff}) the following final conditions (where we denote by $A(u) := \int^u_0 \alpha(y)\ud y$ and set $\phi(X_s) := \alpha^2(X_s)/2 + \alpha'(X_s)/2$),
\begin{cond}[$\Phi$] \label{cond:phi}
There exists a constant $\Phi > -\infty$ such that $\Phi \leq \phi$.
\end{cond}
\begin{cond}[$\varkappa$] \label{cond:cjea}
We have that $\exists\, \varkappa < \infty$ such that,
\begin{align}
\dfrac{\lambda(X_{\psi_i-})\cdot f_\nu\left(X_{\psi_i}; X_{\psi_i-}\right)\cdot e^{-\left[A(X_{\psi_i})-A(X_{\psi_{i}-})\right]}}{\Lambda \cdot f_\delta\left(X_{\psi_i};X_{\psi_i-}\right)} \leq \varkappa. \nonumber
\end{align}
\end{cond}

First considering the \rnd of $\mathbbm{Q}^{x}_{0,T}$ with respect to $\mathbbm{W}^{x}_{0,T}$ we have,
\begin{res}[Unconditioned \rnd \cite{BK:ASCJD}] \label{res:radon}
Under Conditions \ref{cond:exis}--\ref{cond:grow} and \ref{cond:cjea}, the \rnd of $\mathbbm{Q}^x_{0,T}$ with respect to $\mathbbm{W}^x_{0,T}$ exists and is given by Girsanov's formula as follows,
\begin{align}
\dfrac{\ud \mathbbm{Q}^x_{0,T}}{\ud \mathbbm{W}^x_{0,T}}(X) & = \exp\left\{A(X_T)-A(x)-\int^T_0 \phi(X_{s-})\ud s \right\} \cdot \exp\left\{-\int^T_0\left[\lambda(X_{s-}) -\Lambda\right]\ud s\right\}\nonumber\\
& \quad  \cdot \prod^{N_T}_{i=1}\left[\dfrac{\lambda(X_{\psi_i-})\cdot f_\nu\left(X_{\psi_i}; X_{\psi_i-}\right)\cdot e^{-\left[A(X_{\psi_i})-A(X_{\psi_{i}-})\right]}}{\Lambda \cdot f_\delta\left(X_{\psi_i}; X_{\psi_i-}\right)}\right]. \nonumber
\end{align}
In the particular case where we have a diffusion (where $\lambda=\Lambda= 0$), we have,
\begin{align}
\dfrac{\ud \mathbbm{Q}^x_{0,T}}{\ud \mathbbm{W}^x_{0,T}}(X) = \exp\left\{A(X_T)-A(x)-\int^T_0 \phi(X_s)\ud s\right\}.\nonumber
\end{align}
\end{res}

Now considering the \rnd of $\mathbbm{Q}^{x,y}_{0,T}$ with respect to $\mathbbm{W}^{x,y}_{0,T}$, we further denote by $p_{T}(x,y) := \mathbbm{P}_{\mathbbm{Q}^x_{0,T}}(X_T \in \ud y\,|\, X_0 =x)/ \ud y$ and $w_{T}(x,y) := \mathbbm{P}_{\mathbbm{W}^x_{0,T}}(X_T \in \ud y\,|\, X_0 =x)/ \ud y$ as the transition densities of (\ref{eq:dXt}) and (\ref{eq:w radon diff}) respectively over the interval of length $T$ initialised at $X_0=x$.
\begin{res}[Conditioned \rnd \cite{S:DF86}] \label{res:transition}
Following directly from \resref{res:radon} we have,
\begin{align}
\dfrac{\ud \mathbbm{Q}^{x,y}_{0,T}}{\ud \mathbbm{W}^{x,y}_{0,T}}(X) & = \dfrac{w_T(x,y)}{p_T(x,y)}\cdot\dfrac{\ud \mathbbm{Q}^{x}_{0,T}}{\ud \mathbbm{W}^{x}_{0,T}}(X),\nonumber
\end{align}
with transition density of the following form (by taking expectations with respect to $\mathbbm{W}^{x,y}_{0,T}$),
\begin{align}
p_{T}(x,y)
& = w_{T}(x,y)\cdot\mathbbm{E}_{\mathbbm{W}^{x,y}_{0,T}}\left[\dfrac{\ud \mathbbm{Q}^x_{0,T}}{\ud \mathbbm{W}^x_{0,T}}(X)\right]. \nonumber
\end{align}
\end{res}

Throughout this paper we rely on the fact that upon simulating a path space layer (see \defnref{defn:layer}) then $\forall s \in[0,T]$ $\phi(X_s)$ is bounded, however this follows directly from the following result,
\begin{res}[Local Boundedness] \label{res:compact}
By \conref{cond:cont}, $\alpha$ and $\alpha'$ are bounded on compact sets. In particular, suppose $\exists\,\ell, \upsilon \in \mathbbm{R}$ such that $\forall$ $t\in[0,T]$, $X_t(\omega)\in[\ell,\upsilon]$ $\exists\,L_X:=L\left(X(\omega)\right)\in\mathbbm{R},U_X:=U\left(X(\omega)\right)\in\mathbbm{R}$ such that $\forall$ $t\in[0,T]$, $\phi\left(X_t(\omega)\right)\in[L_X,U_X]$.
\end{res}

\section[EXACT SIMULATION OF CONDITIONED DIFFUSIONS]{EXACT SIMULATION OF CONDITIONED DIFFUSIONS} \label{s:cauea} 
In this section we outline exact algorithms to simulate sample path skeletons of diffusion bridges (under Conditions \ref{cond:exis}--\ref{cond:phi} and following the Lamperti transform (\resref{res:lamp})) which can be represented as the solution to the following SDE,
\begin{align}
\ud X_t 
& = \alpha(X_t) \ud t + \ud W_t, \quad X_0 = x \in\mathbbm{R},\, X_T = y \in\mathbbm{R},\,t\in[0,T]. \label{eq:conditioned diffusion2}
\end{align}
We present two separate exact algorithms to simulate conditioned diffusion sample path skeletons -- the \textit{Conditioned Unbounded Exact Algorithm (CUEA)} and the \textit{Conditioned Adaptive Unbounded Exact Algorithm (CAUEA)} (which is a Rao-Blackwellisation of the CUEA requiring less simulation of the sample path). The methodology developed in this section is a direct extension of that developed for unconditioned diffusions in \cite{B:PJR15} (termed the \textit{Unbounded Exact Algorithm} and  \textit{Adaptive Unbounded Exact Algorithm} respectively), but also serves to introduce the key ideas for when we consider the non-trivial extension to the simulation of jump diffusion bridge sample path skeletons in \secref{s:caujea}.

Exact algorithms are a class of rejection samplers operating on diffusion path space (introduced by \cite{AAP:BR05}) in which finite dimensional subsets of sample paths are drawn from $\mathbbm{Q}^{x,y}_{0,T}$ (recall, the measure induced by (\ref{eq:conditioned diffusion2})) by means of simulating finite dimensional subsets of sample paths from an (easy to simulate) equivalent measure with bounded \rnd. As established in \secref{s:preliminaries}, $\mathbbm{W}^{x,y}_{0,T}$ is such an equivalent measure (Brownian motion measure, from which finite dimensional subsets of sample paths can be drawn without error (see \cite[\textsection 2.8]{PhD:P13})). Proceeding as in standard rejection sampling, if we draw $X\sim\mathbbm{W}^{x,y}_{0,T}$ and accept the sample path $(I=1)$ with probability ${{P}}_{\mathbbm{W}^{x,y}_{0,T}}(X) := \frac{1}{M}\frac{\ud \mathbbm{Q}^{x,y}_{0,T}}{\ud \mathbbm{W}^{x,y}_{0,T}}(X)\in[0,1]$ then $(X|I=1)\sim\mathbbm{Q}^{x,y}_{0,T}$. Now, considering the form of the acceptance probability we have,
\begin{theorem}[Conditioned Exact Algorithm Acceptance Probability I] \label{thm:acceptprob1}
\noindent $\mathbbm{Q}^{x,y}_{0,T}$ is equivalent to $\mathbbm{W}^{x,y}_{0,T}$ with \rnd{}:
\begin{align}
\dfrac{\ud \mathbbm{Q}^{x,y}_{0,T}}{\ud \mathbbm{W}^{x,y}_{0,T}}(X)  &\propto \exp\left\{-\int^T_0 \phi(X_s)\ud s\right\} \in \left[0,e^{-\Phi T}\right], \label{eq:acceptprob}
\end{align}
and so we have that,
\begin{align}
{{P}}_{\mathbbm{W}^{x,y}_{0,T}}(X) = e^{\Phi T}\cdot \exp\left\{-\int^T_0 \phi(X_s)\ud s\right\}  \in [0,1]. \label{eq:waccprob}
\end{align}
\proof
LHS of (\ref{eq:acceptprob}) from Results \ref{res:radon} and \ref{res:transition}. RHS of (\ref{eq:acceptprob}) from \conref{cond:phi}. (\ref{eq:waccprob})  rearranged from (\ref{eq:acceptprob}).
\endproof
\end{theorem}

As remarked in \secref{s:introduction}, it isn't possible to simulate entire diffusion sample paths (they are infinite dimensional) and so it isn't possible to evaluate the integral in (\ref{eq:waccprob}). However, it was noted in \cite{AAP:BR05,B:BPR06,MCAP:BPR08} (and summarised in \algref{alg:ea}) that by first simulating an auxiliary random variable $F\sim\mathbbm{F}$, an unbiased estimator of (\ref{eq:acceptprob}) can be constructed and evaluated without having to simulate the entire sample path (i.e. $F$ informs us as to which parts of the sample path to simulate (denoted $\fin{X}$)).  The remainder of the sample path (denoted $\rem{X}:=X\setminus \fin{X} $) can be simulated as required \textit{after} acceptance (hence the asterisk in \algstref{alg:ea}{alg:ea:inf}) conditional on its skeleton (composed of $F$, $X_0=x$, $X_T=y$ and $\fin{X}$).  
\begin{algorithm}[h]
	\caption{Exact Algorithm for Conditioned Diffusions.} \label{alg:ea}
	\begin{enumerate}
	\item Simulate $F\sim\mathbbm{F}$. \label{alg:ea:var}
	\item Simulate $\fin{X}\sim \left.\mathbbm{W}^{x,y}_{0,T}\mvbar F\right.$. \label{alg:ea:cond}
	\item With probability ${{P}}_{\mathbbm{W}^{x,y}_{0,T}| F}\left(X\right)$ accept, else reject and return to \stepref{alg:ea:var}. \label{alg:ea:prob}\\
	\vspace{-0.15cm}\hrule\vspace{0.15cm}
	\item \textit{* Simulate $\rem{X}\sim \left.\mathbbm{W}^{x,y}_{0,T}\mvbar(\fin{X},F)\right.$.} \label{alg:ea:inf}
	\vspace{-0.35cm}
	\end{enumerate}
\end{algorithm}

Now we consider how to construct a suitable finite dimensional random variable $F\sim\mathbbm{F}$ (while ensuring we satisfy Principles \ref{prin:layer}--\ref{prin:rest}). As noted in \secref{s:preliminaries}, to simulate a sample path skeleton we will typically require a path space \textit{layer}. This is due to the fact that the method employed to construct $\mathbbm{F}$ requires upper and lower bounds for $\phi(X_{[0,T]})$ which, as a consequence of \resref{res:compact}, is provided by a path space layer ($U_X\in\mathbbm{R}$ and $L_X\in\mathbbm{R}$ respectively). As such the first step in simulating $\mathbbm{F}$ is to partition the path space of $\mathbbm{W}^{x,y}_{0,T}$ into disjoint layers and simulate the layer to which our proposal sample path belongs (see \prinref{prin:layer}, denoting $R:=R(X)\sim\mathcal{R}$ as the simulated layer). As such we have for all test functions $H\in\mathcal{C}_\text{b}$,
\begin{align}
\mathbbm{E}_{\mathbbm{W}^{x,y}_{0,T}}\left[P_{\mathbbm{W}^{x,y}_{0,T}}(X)\cdot H(X)\right] &= \mathbbm{E}_\mathcal{R}\mathbbm{E}_{\left.\mathbbm{W}^{x,y}_{0,T}|R\right.}\left[{{P}}_{\mathbbm{W}^{x,y}_{0,T}}(X)\cdot H(X)\right]. \nonumber
\end{align}
Conditional on the simulated layer we can represent our acceptance probability as follows,
\begin{align}
{{P}}_{\mathbbm{W}^{x,y}_{0,T}}(X) & = e^{-(L_X\!-\!\Phi)T} \cdot \exp\left\{-\int^T_0\!\left(\phi(X_s)\!-\!L_X\right)\ud s\right\} =: e^{-(L_X-\Phi)T}\cdot {{\tilde{P}}}_{\mathbbm{W}^{x,y}_{0,T}|R}(X),  \label{eq:pois ap ex}
\end{align}
noting that,
\begin{align}
{{\tilde{P}}}_{\mathbbm{W}^{x,y}_{0,T}|R}(X)\in\left[e^{-\left(U_X-L_X\right)T},1\right] \subseteq (0,1]. \label{eq:pois ap}
\end{align}

\begin{figure}[h]
\centering
\includegraphics[width=0.5\textwidth]{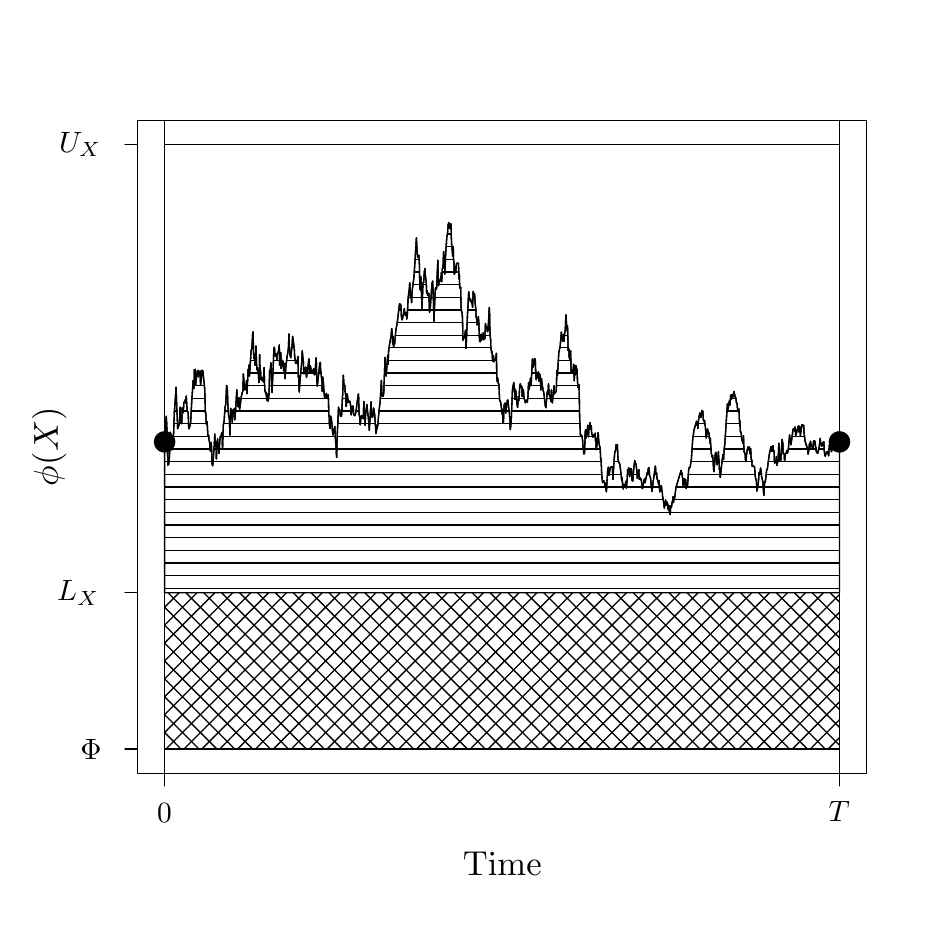}
\caption{Example trajectory of $\phi(X)$ where $X\sim\left.\mathbbm{W}^{x,y}_{0,T}\mvbar R(X)\right.$.} \label{fig:cphil}
\end{figure}

As noted in \cite{B:BPR06} and with the aid of \figref{fig:cphil}, ${{P}}_{\mathbbm{W}^{x,y}_{0,T}}(X)$ is precisely the probability a Poisson process of intensity $1$ on the graph $\mathcal{G}_A:=\{(x,y)\in[0,T]\times[\Phi,\infty):y\leq\phi(x)\}$ contains no points. This process can be simulated using a Poisson thinning argument, by means of simulating a Poisson process of intensity $1$ on the larger graph $\mathcal{G}_P:=[0,T]\times[\Phi,U_X]\supseteq\mathcal{G}_A$ (which is trivial), computing $\phi(X)$ at a finite collection of time points and then determining whether or not any of the points lie in $\mathcal{G}_A$. With reference to (\ref{eq:pois ap ex}) and as noted in \cite{B:PJR15} (and in the related later equivalent construction of \cite{JAP:D14}), this approach to simulating an event of probability ${{P}}_{\mathbbm{W}^{x,y}_{0,T}}(X)$ can be made computationally more efficient be deploying an \textit{accelerated rejection} strategy, in which the sample path is first rejected with probability $1-e^{-(L_X-\Phi)T}$ ($\in[0,1)$, the crosshatched region in \figref{fig:cphil}) and then, conditional on not having been rejected, acceptance is determined by simulating an additional event of probability ${{\tilde{P}}}_{\mathbbm{W}^{x,y}_{0,T}|R}(X)$ (the vertically hatched region in \figref{fig:cphil} which can be simulated as per ${{P}}_{\mathbbm{W}^{x,y}_{0,T}}(X)$, but with the alternate graphs of $\mathcal{G}_A:=\{(x,y)\in[0,T]\times[L_X,\infty):y\leq\phi(x)\}$ and $\mathcal{G}_P:=[0,T]\times[L_X,U_X]$). The critical observation in these approaches is that the acceptance probability can be evaluated using only a finite dimensional realisation of the sample path, $\fin{X}$. The above argument is stated more formally in \thmref{thm:acceptprob2}, with \algref{alg:cuea} detailing how to implement this strategy to simulate sample path skeletons.
\begin{theorem}[Conditioned Exact Algorithm Acceptance Probability II \protect{\cite[\textsection 3.1]{B:PJR15}}] \label{thm:acceptprob2}
Letting $\mathbbm{K}_R$ be the law of $\kappa \sim \text{Poi}((U_X-L_X) T)$, $\mathbbm{U}\kappa$ the distribution of $(\xi_1,\ldots{},\xi_\kappa)\overset{\text{iid}}{\sim} \U[0,T]$ we have,
\begin{align}
{{P}}_{\mathbbm{W}^{x,y}_{0,T}}(X)
 = e^{-(L_X-\Phi)T}\cdot\mathbbm{E}_{\mathbbm{K}_{R}}\left[\mathbbm{E}_{\mathbbm{U}\kappa}\left[\prod^\kappa_{i=1}\left(\dfrac{U_X\!-\!\phi(X_{\xi_i})}{U_X\!-\!L_X}\right)\,\mvbar\,X\right]\mvbar\,X\right]. \nonumber
\end{align}
\end{theorem}
\begin{algorithm}[h]
	\caption{Conditioned Unbounded Exact Algorithm (CUEA).} \label{alg:cuea}
	\begin{enumerate}
	\item Simulate layer information $R\sim\mathcal{R}$ as per \cite[\textsection 7.1]{B:PJR15}. \label{alg:cuea:start}\label{alg:cuea:layer}
	\item With probability $\left(1-\exp\left\{-(L_X-\Phi)T\right\}\right)$ reject path and return to \stepref{alg:cuea:start}. \label{alg:cuea:pre}
	\item Simulate skeleton points $\left.\left(X_{\xi_1},\ldots{},X_{\xi_\kappa}\right)\mvbar R\right.$,  \label{alg:cuea:lbb}
	\begin{enumerate}
		\item Simulate $\kappa\sim \text{Poi}\big((U_X-L_X) T\big)$ and skeleton times $\xi_1,\ldots{},\xi_\kappa \overset{\text{iid}}{\sim} \U[0,T]$. \label{alg:cuea:poisskel}
		\item Simulate sample path at skeleton times $X_{\xi_1},\ldots{},X_{\xi_\kappa}\sim\left.\mathbbm{W}^{x,y}_{0,T}\mvbar R\right.$ as per \cite[\textsection 7.1]{B:PJR15}.
	\end{enumerate}
	\item With probability $\prod^\kappa_{i=1}\left[\left(U_X-\phi(X_{\xi_i})\right)/\left(U_X-L_X\right)\right]$, accept path, else reject and return to \stepref{alg:cuea:start}. \label{alg:cuea:acpr}\\
	\vspace{-0.25cm}\hrule\vspace{0.15cm}
	\item \textit{* Simulate $\rem{X}\sim\left.\left(\otimes^{\kappa+1}_{i=1} \mathbbm{W}^{X_{\xi_{i-1}},X_{\xi_i}}_{\xi_{i-1},\xi_i}\right)\mvbar R\right.$} as per \cite[\textsection 3.1]{B:PJR15}. \label{alg:cuea:inf}
	\vspace{-0.35cm}	
	\end{enumerate}
\end{algorithm}

The computational cost of the CUEA is intrinsically linked to the area of the graph $\mathcal{G}_P$, and so we naturally want to choose or construct the graph $\mathcal{G}_P$ to occupy as small an area as possible. It was noted in \cite[\textsection 3.2]{B:PJR15} that \algstref{alg:cuea}{alg:cuea:poisskel} could be equivalently performed by means of simulating exponential random variables. We could for instance set $\xi_0=0$ and iteratively set $\xi_i=\xi_{i-1}+\zeta_i$ where $\zeta_i \sim \text{Exp}(U_X-L_X)$ while $\sum_i \zeta_i \leq T$, or in any other convenient order provided we have coverage of the interval $[0,T]$. The key idea in \cite[\textsection 3.2]{B:PJR15} is to use this iterative simulation of the sample path to construct an \textit{Adaptive Exact Algorithm} in which we find refined upper and lower bounds for segments of $\phi(X_{[0,T]})$, and hence accelerate the acceptance or rejection of the sample path (in essence find a smaller graph $\mathcal{G}_P$ to conduct the remainder of the simulation). This approach is well suited to simulating conditioned diffusion sample paths as, as noted in \secref{s:introduction}, over long time intervals the computational cost for employing an exact algorithm for conditioned diffusions can be infeasible (the bounds on the path space layer are less tight and hence the graph $\mathcal{G}_P$ is larger).

As discussed in \cite[\textsection 3.2]{B:PJR15}, the most computationally efficient order of simulating the exponential random variables is iteratively emanating from the centre of uncovered intervals (where there is the opportunity to learn most about the extent to which the sample path oscillates). In particular, beginning at the interval mid-point ($T/2$), we can find the skeletal point closest to the mid-point by simulating $\tau\sim \text{Exp}(2[U_X-L_X])$ and setting the skeletal point ($\xi'$) to be with equal probability either $T/2-\tau$ or $T/2+\tau$. Halting our simulation of (\ref{eq:pois ap}) at this point we arrive at (\ref{eq:auea con dec}) where we have decomposed our acceptance probability into the product of three probabilities associated with three disjoint sub-intervals (conditional on $\xi'\in[0,T]$, we have $[0,T]=[0,T/2-\tau]\uplus[T/2-\tau,T/2+\tau]\uplus[T/2+\tau,T]$). If we consider the evaluation of each successively we need only continue to the next (and expend computation) conditional on the previous being accepted (i.e. we have an accelerated rejection strategy). We begin by evaluating the computationally cheap expectation in (\ref{eq:auea con dec}) (which is with respect to $u\sim \U[0,1]$), before proceeding to the acceptance probabilities for the left and right sub-intervals, each of which has the same form as (\ref{eq:pois ap}).
\begin{align}
\label{eq:auea con dec} {\tilde{P}}_{\mathbbm{W}^{x,y}_{0,T}|R,X_{\xi'}}(X)
& = \left\{\!\!\!\!\! 
\begin{array}{l l}
\begin{array}{l}
\mathbbm{E}\left(\mathbbm{1}\left[u\leq \dfrac{U_X-\phi(X_{\xi'})}{U_X-L_X}\right]\mvbar X_{\xi'}\right) \\
\quad \cdot \exp\left\{-\int^{T/2-\tau}_0\left[\phi(X_s)-L_X\right]\ud s -\int^T_{T/2+\tau} \left[\phi(X_s)-L_X\right]\ud s\right\}
\end{array}, & \!\!\text{if $\xi' \in [0,T]$}, \\
\begin{array}{l}1\end{array}, & \!\!\text{if $\xi' \notin [0,T]$}. \\
 \end{array}\right.
\end{align}
Considering in isolation the acceptance probability corresponding to the interval $[0,T/2-\tau]$ in (\ref{eq:auea con dec}), we can now find new layer information ($R^{[0,{\xi'}]}_X$) which more tightly bounds the sample path  and so find tighter bounds for $\phi(X_{[0,\xi']})$ (denoted $U_X^{[0,\xi']}$ and $L_X^{[0,\xi']}$). As such the acceptance probability can be re-written,
\begin{align}
 \exp\Big\{-\!\int^{\frac{T}{2}-\tau}_0\!\![\phi(X_s)\!-\!L_X]\!\ud s\Big\}
& = \exp\Big\{\!-\!(L^{[0,\xi']}_X\!-\!L_X)\cdot(T/2\!-\!\tau)\!\Big\}\cdot \exp\Big\{-\!\int^{\frac{T}{2}-\tau}_0\!\![\phi(X_s)\!-\!L^{[0,\xi']}_X]\!\ud s\Big\}. \label{eq:lhs dec}
\end{align}
The form of (\ref{eq:lhs dec}) now coincides with (\ref{eq:pois ap ex}) and so can be evaluated using the same procedure outlined above. Iterating this procedure until the entire sample path is accepted or rejected results in the \textit{Conditioned Adaptive Unbounded Exact Algorithm (CAUEA)} presented in \algref{alg:cauea}. In  \algref{alg:cauea} we use the following notation: $\Pi$ denotes the set comprising information required to evaluate the acceptance probability for each interval still to be estimated, $\Pi:=\left\{\Pi(i)\right\}^{|\Pi|}_{i=1}$. Each $\Pi(i)$ comprises information regarding the time interval it applies to $\left[s(\Pi(i)),t(\Pi(i))\right]$, the sample path at known points at either side of this interval ($x(\Pi(i)):=X^{\Pi(i)}_{\bar{s}}$, $y(\Pi(i)):=X^{\Pi(i)}_{\bar{t}}$) and the associated layer ($R^{\Pi(i)}_X$) and induced bounds on $\phi$ ($U^{\Pi(i)}_X$ and $L^{\Pi(i)}_X$), noting that $\bar{s}\leq s<t\leq \bar{t}$. We further denote $2m(\Pi(i)) := [s(\Pi(i))+t(\Pi(i))]$, $2d(\Pi(i)) := [t(\Pi(i))-s(\Pi(i))]$.
\begin{algorithm}[h]
	\caption{Conditioned Adaptive Unbounded Exact Algorithm (CAUEA).} \label{alg:cauea}
	\begin{enumerate}
	\item Simulate layer information $R_X\sim\mathcal{R}$  as per \cite[\textsection 8.1]{B:PJR15}, setting $\Pi := \left\{\Xi\right\} := \left\{\left\{[0,T], X_0, X_T, R_X\right\}\right\}$ and $\kappa=0$. \label{alg:cauea:layer}  \label{alg:cauea:start}
	\item With probability $\left(1-\exp\left\{-(L_X-\Phi)T\right\}\right)$ reject path and return to \stepref{alg:cauea:start}. \label{alg:cauea:prelim}
	\item Set $\Xi=\Pi(1)$. \label{alg:cauea:loop}
	\item Simulate $\tau\sim \text{Exp}\left(2[U^{\Xi}_X-L^{\Xi}_X]\right)$. If $\tau > d(\Xi)$ then set $\Pi:=\Pi\setminus\Xi$ else,
	\begin{enumerate}
	\item Set $\kappa=\kappa+1$ and with probability $1/2$ set $\xi'_\kappa=m(\Xi)-\tau$ else $\xi'_\kappa=m(\Xi)+\tau$.
	\item Simulate $X_{\xi'_\kappa}\sim\left.\mathbbm{W}^{x(\Xi), y(\Xi)}_{\bar{s}(\Xi),\bar{t}(\Xi)}\,|\, R^{\Xi}_X\right.$ as per \cite[\textsection 8.2]{B:PJR15}. \label{alg:cauea:conlayer}
	\item With probability $\left(1-[U^{\Xi}_X-\phi(X_{\xi'_\kappa})]/[U^{\Xi}_X-L^{\Xi}_X]\right)$ reject sample path and return to \stepref{alg:cauea:start}.
	\item Simulate new layer information $R_X^{[\bar{s}(\Xi),\xi'_\kappa]}$ and $R_X^{[\xi'_\kappa,\bar{t}(\Xi)]}$ conditional on $R^{\Xi}_X$ as per \cite[\textsection 8.3 \& \textsection 8.4]{B:PJR15}. \label{alg:cauea:simlayer}
	\item With probability $\Big(1-\exp\Big\{-\Big[L^{[\bar{s}(\Xi),\xi'_\kappa]}_X+L^{[\xi'_\kappa,\bar{t}(\Xi)]}_X-2L^{\Xi}_X\Big]\cdot \left[d(\Xi)-\tau\right]\Big\}\Big)$ reject sample path and return to \stepref{alg:cauea:start}.
	\item Set  $\Pi:=\Pi\bigcup\Big\{\!\!\left[s(\Xi),m(\Xi)\!-\!\tau\right], X^{\Xi}_{\bar{s}}, X_{\xi'_\kappa}, R^{[\bar{s}(\Xi),\xi'_\kappa]}_X\!\Big\}\bigcup\Big\{\!\!\left[m(\Xi)\!+\!\tau,t(\Xi)\right], X_{\xi'_\kappa}, X^\Xi_{\bar{t}}, R^{[\xi'_\kappa,\bar{t}(\Xi)]}_X\!\Big\}\!\setminus\!\Xi$.
	\end{enumerate}
	\item If $|\Xi|\neq0$ return to \stepref{alg:cauea:loop}.
	\item Define skeletal points $\xi_1,\ldots{},\xi_\kappa$ as the order statistics of the set $\left\{\xi'_1,\ldots{},\xi'_\kappa\right\}$.\\
	\vspace{-0.25cm}\hrule\vspace{0.15cm}
	\item \textit{* Simulate $\rem{X} \sim \left(\otimes^{\kappa+1}_{i=1} \mathbbm{W}^{X_{\xi_{i-1}},X_{\xi_i}}_{\xi_{i-1},\xi_i}\mvbar R^{[\xi_{i-1},\xi_i]}_X\right)$}  as per \cite[\textsection 8.5]{B:PJR15}. \label{alg:cauea:inf}
	\vspace{-0.35cm}
	\end{enumerate}
\end{algorithm}

\begin{figure}[h] 
\centering
\subfigure[Example skeleton output from the CUEA (\algref{alg:cuea}), $\mathcal{S}_{\text{CUEA}}\left(X\right)$, overlaid with two possible sample path trajectories consistent with the skeleton.\label{fig:ill:cuea}]{
\includegraphics[width=0.305\textwidth]{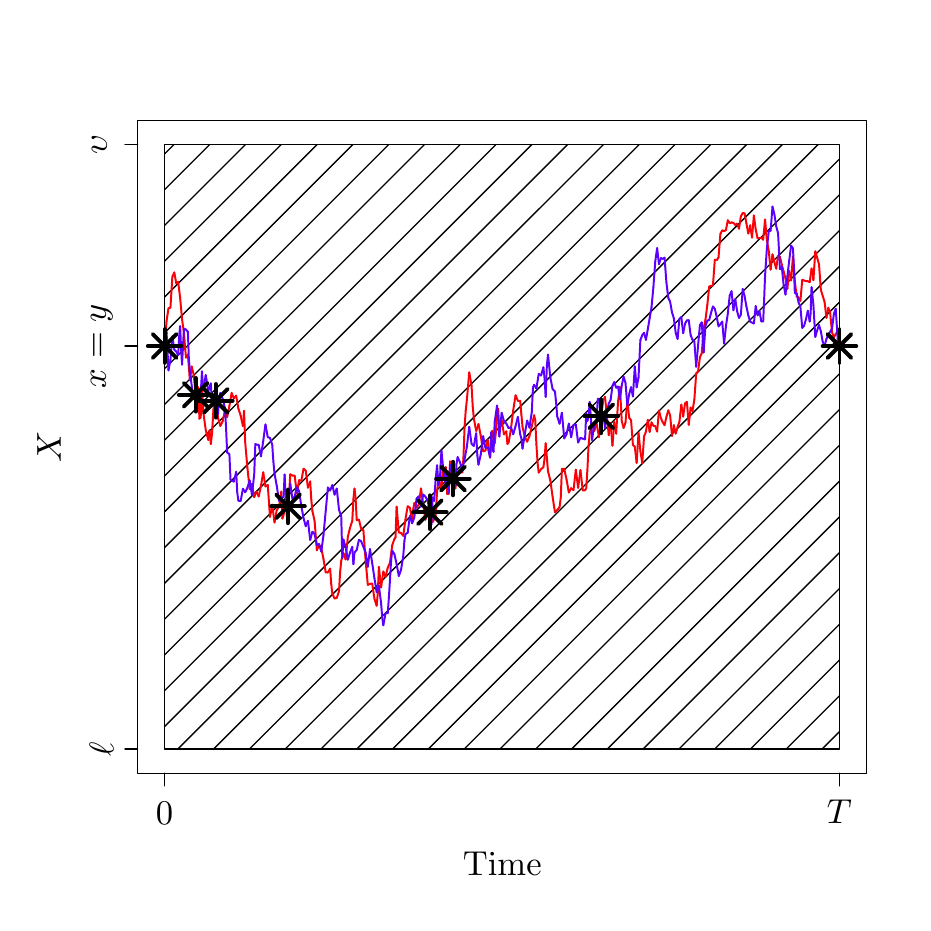}} \hspace*{0.25cm}
\subfigure[Example skeleton output from the CAUEA (\algref{alg:cauea}), $\mathcal{S}_{\text{CAUEA}}\left(X\right)$, overlaid with two possible sample path trajectories consistent with the skeleton. \label{fig:ill:cauea}]{
\includegraphics[width=0.305\textwidth]{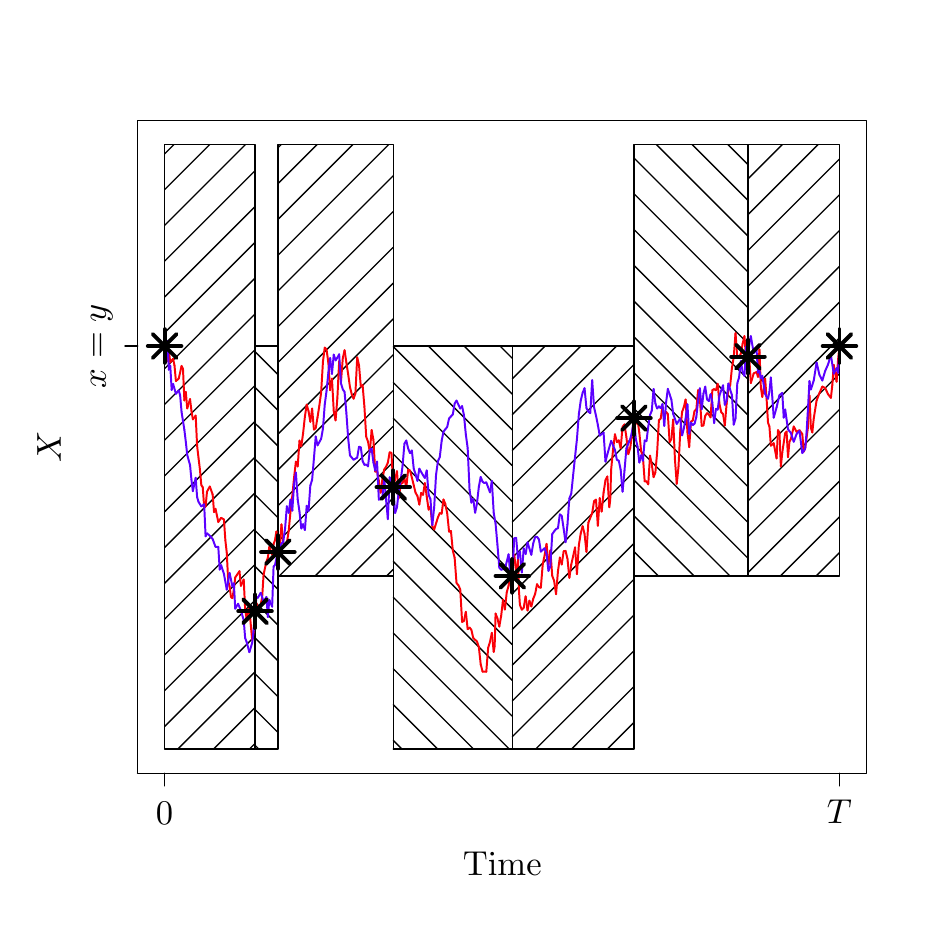}} \hspace*{0.25cm}
\subfigure[Example skeleton output from the CAUJEA (\algref{alg:caujea}), $\mathcal{S}_{\text{CAUJEA}}\left(X\right)$, overlaid with two possible sample path trajectories consistent with the skeleton. \label{fig:ill:caujea}]{
\includegraphics[width=0.305\textwidth]{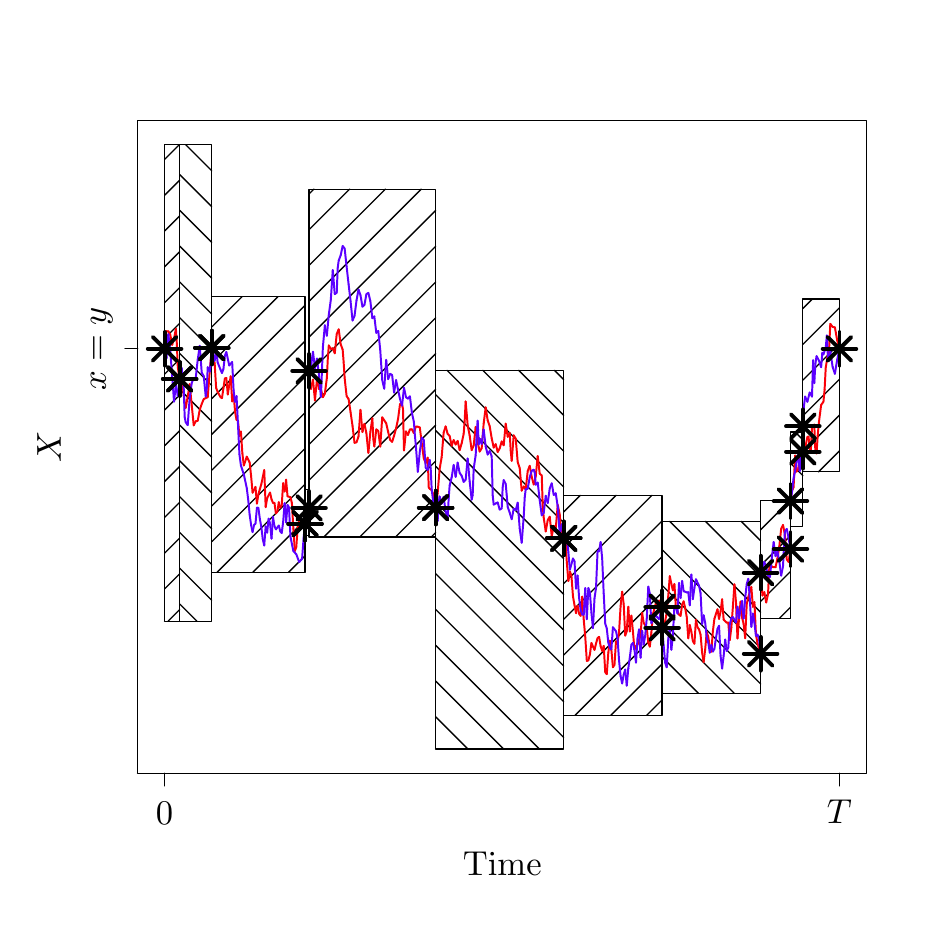}} 
\caption{Comparison of the CUEA, CAUEA and CAUJEA skeleton output. Hatched regions indicate layer information, whereas the asterisks indicate skeletal points.} \label{fig:ill:cueacauea}
\end{figure}

Accepted sample path skeletons simulated under both the CUEA and CAUEA are composed of given terminal points, skeletal points and layer information and have a form as shown in (\ref{eq:cdiffskel}). Both approaches satisfy Principles \ref{prin:layer}--\ref{prin:rest} (although, the CUEA requires augmentation with additional layer information as per \cite[\textsection 3.1]{B:PJR15}). In Figures \ref{fig:ill:cuea} and \ref{fig:ill:cauea} we present illustrative examples of accepted sample path skeletons under the two approaches.
\begin{align}
\mathcal{S}_\text{CUEA}\left(X\right) :=\left\{\left(\xi_i,X_{\xi_i} \right)^{\kappa+1}_{i=0}, R\right\}, \quad \mathcal{S}_\text{CAUEA}\left(X\right) := \left\{\left(\xi_i,X_{\xi_i}\right)^{\kappa+1}_{i=0}, \left(R^{[\xi_{i-1},\xi_i]}_X\right)^{\kappa+1}_{i=1}\right\}.  \label{eq:cdiffskel}
\end{align}

\TMResetAll
\section[EXACT SIMULATION OF CONDITIONED JUMP DIFFUSIONS]{EXACT SIMULATION OF CONDITIONED JUMP DIFFUSIONS} \label{s:caujea} 
In this section we extend the methodology of \secref{s:cauea}, outlining how to simulate sample path skeletons of conditioned jump diffusions (under Conditions \ref{cond:exis}--\ref{cond:cjea} and following the Lamperti transform (\resref{res:lamp})) which can be represented as the solution to the following SDE (denoting $X_{t-} := \lim_{s \uparrow t }X_s$),
\begin{align} 
\ud X_t & = \alpha(X\tm) \ud t + \ud W_t + \ud J^{\lambda,\nu}_t, \quad\quad X_0 = x \in\mathbbm{R},\, X_T=y \in\mathbbm{R},\, t\in[0,T]. \label{eq:cjdiff}
\end{align}
The approach we take in this section in constructing our exact algorithm is based upon the recent methodology developed in \cite{MCAP:GR13}. However, we reformulate the exact algorithm presented in \cite{MCAP:GR13} to ensure that upon accepting a sample path skeleton then it is possible to simulate the sample path at further finite collections of time points (i.e. it satisfies Principles \ref{prin:layer}--\ref{prin:rest}) and in order to employ accelerated rejection strategies to reduce the computational cost of simulation.

The rejection sampling construction of \secref{s:cauea} to simulate sample skeletons from $\mathbbm{Q}^{x,y}_{0,T}$ cannot be directly employed in the case of conditioned jump diffusions (\ref{eq:cjdiff}) with $\mathbbm{W}^{x,y}_{0,T}$ as the proposal measure, as it is not possible to simulate a compound Poisson process conditioned to hit a specified end point. The key contribution of \cite{MCAP:GR13} was to note that an alternate equivalent measure (denoted $\mathbbm{G}^{x,y}_{0,T}$) can be constructed to ensure the end point is hit. In particular, if a compound Poisson process is simulated first ($J_{[0,T]}$) then, to ensure the end point is hit ($X_T=y$), a Brownian bridge conditioned to start at $X'_0:=x'=x$ and end at $X'_T:=y'=y-J_T$ can be used as the continuous component in the proposal sample path. Considering the superposition of the compound Poisson process sample path and the Brownian bridge sample path ($X_t=J_t + X'_t$), then the resulting sample path starts and ends at the desired points ($X_0=x$ and $X_T=y$). More formally $\mathbbm{G}^{x,y}_{0,T}$ is the measure induced by the following SDE,
\begin{align}
\ud X_t & = \ud Z_t + \ud J^{\Lambda,\delta}_t,\quad\quad X_0 = x \in\mathbbm{R},\,X_T=y \in\mathbbm{R},\, t\in[0,T], \label{eq:cjeaprop2.1}
\end{align}
where $Z\sim\mathbbm{\overline{W}}^{x,y'}_{0,T}$ (where $\mathbbm{\overline{W}}^{x,y'}_{0,T}$ is Brownian bridge measure starting at $Z_0=x$ and ending at $Z_T=y'=y-J_T$).

Proceeding as in \secref{s:cauea}, we require the \rnd of $\mathbbm{Q}^{x,y}_{0,T}$ with respect to $\mathbbm{G}^{x,y}_{0,T}$.
\begin{theorem}[\rnd for conditioned jump diffusions \protect{\cite[Lemma 2]{MCAP:GR13} \cite[Thm. 5.4.1]{PhD:P13}}] \label{thm:rndjea}
\noindent $\mathbbm{Q}^{x,y}_{0,T}$ is equivalent to $\mathbbm{G}^{x,y}_{0,T}$ with \rnd:
\begin{align}
\dfrac{\ud \mathbbm{Q}^{x,y}_{0,T}}{\ud \mathbbm{G}^{x,y}_{0,T}}(X) 
& \propto \underbrace{\exp\left\{-\dfrac{1}{2}\dfrac{\left(y-J_T-x\right)^2}{T}\right\}}_{\leq 1}\cdot\underbrace{\exp\left\{-\int^T_0 \phi(X_{s-})\ud s\right\}}_{\leq\,\exp\left\{-\Phi T\right\}} \cdot \underbrace{\exp\left\{-\int^T_0\left[\lambda(X_{s-}) -\Lambda\right]\ud s\right\}}_{\leq\exp\left\{\Lambda T\right\}} \nonumber\\
& \quad\quad\quad  \cdot \underbrace{\prod^{N_T}_{i=1}\dfrac{\lambda(X_{\psi_i-})\cdot f_\nu\left(X_{\psi_i}; X_{\psi_i-}\right)\cdot \exp\left\{-\left[A(X_{\psi_i})-A(X_{\psi_{i}-})\right]\right\}}{\Lambda \cdot f_\delta\left(X_{\psi_i};X_{\psi_i-}\right)}}_{\leq\,\varkappa^{N_T}}. \label{eq:qgrnd}
\end{align}
\end{theorem}
\noindent Following our exact algorithm construction of \secref{s:cauea}, if we simply draw $X\sim\mathbbm{G}^{x,y}_{0,T}$ and accept the sample path $(I=1)$ with probability ${{P}}_{\mathbbm{G}^{x,y}_{0,T}}(X) := \frac{1}{M}\frac{\ud \mathbbm{Q}^{x,y}_{0,T}}{\ud \mathbbm{G}^{x,y}_{0,T}}(X)\in[0,1]$, then we have that $(X|I=1)\sim\mathbbm{Q}^{x,y}_{0,T}$. Considering the form of the acceptance probability (by rearrangement of (\ref{eq:qgrnd})) we have,
\begin{align}
{{P}}_{\mathbbm{G}^{x,y}_{0,T}}(X) 
& := \underbrace{e^{\Phi T} \cdot \exp\left\{-\int^T_0 \left[\phi(X_{s-})+\lambda(X_{s-})\right]\ud s\right\}}_{=:P^{(3)}_{\mathbbm{G}^{x,y}_{0,T}}(X)} \cdot \underbrace{\exp\left\{-\dfrac{1}{2}\dfrac{\left(y-J_T-x\right)^2}{T}\right\}}_{=:P^{(1)}_{\mathbbm{G}^{x,y}_{0,T}}(X)} \nonumber\\
& \quad\quad  \cdot \underbrace{\dfrac{1}{\varkappa^{N_T}} \cdot \prod^{N_T}_{i=1}\dfrac{\lambda(X_{\psi_i-})\cdot f_\nu\left(X_{\psi_i}; X_{\psi_i-}\right)\cdot \exp\left\{-\left[A(X_{\psi_i})-A(X_{\psi_{i}-})\right]\right\}}{\Lambda \cdot f_\delta\left(X_{\psi_i};X_{\psi_i-}\right)}}_{=:P^{(2)}_{\mathbbm{G}^{x,y}_{0,T}}(X)}. \label{eq:cjeaaccprob}
\end{align}
As in \secref{s:cauea}, by first simulating a finite dimensional auxiliary random variable $F\sim\mathbbm{F}$ an unbiased estimator of (\ref{eq:cjeaaccprob}) can be constructed and evaluated without having to simulate the entire sample path (leaving us with a sample path skeleton). In this instance the first step in constructing $\mathbbm{F}$ is to follow our construction of the proposal measure $\mathbbm{G}^{x,y}_{0,T}$ in (\ref{eq:cjeaprop2.1}), and simulate the process $J_{[0,T]}\sim\mathcal{J}$ (where $\mathcal{J}$ is the law of the compound Poisson process component of $\mathbbm{G}^{x,y}_{0,T}$). As such we have for all test functions $H\in\mathcal{C}_\text{b}$,
\begin{align}
\mathbbm{E}_{\mathbbm{G}^{x,y}_{0,T}}\left[P_{\mathbbm{G}^{x,y}_{0,T}}(X)\cdot H(X)\right]
& = \mathbbm{E}_{\mathcal{J}}\mathbbm{E}_{\mathbbm{G}^{x,y}_{0,T}} \left[{{P}}_{\mathbbm{G}^{x,y}_{0,T}}(X)\cdot H(X)\,\mvbar\,J_{[0,T]}\right]. \nonumber
\end{align}
Further denoting by $\mathcal{W}\,|\,\mathcal{J}$ as the law induced by simulating $(X'_{\psi_1},\ldots{},X'_{\psi_{N_T}})\sim\mathbbm{\overline{W}}^{x,y'}_{0,T}$ we have, 
\begin{align}
& \mathbbm{E}_{\mathbbm{G}^{x,y}_{0,T}}\left[P_{\mathbbm{G}^{x,y}_{0,T}}(X)\cdot H(X)\right] 
= \mathbbm{E}_{\mathcal{J}}\mathbbm{E}_{\mathbbm{G}^{x,y}_{0,T}} \left[P^{(1)}_{\mathbbm{G}^{x,y}_{0,T}}(X) \cdot  P^{(2)}_{\mathbbm{G}^{x,y}_{0,T}}(X) \cdot  P^{(3)}_{\mathbbm{G}^{x,y}_{0,T}}(X)\cdot H(X)\,\mvbar\, N_T, \{\psi_i\}^{N_T}_{i=1}, \{\delta_i\}^{N_T}_{i=1} \right] \label{eq:pg1}\\
& = \mathbbm{E}_{\mathcal{J}} \mathbbm{E}_{\mathcal{W\,|\,J}}\mathbbm{E}_{\mathbbm{\overline{W}}^{x,y'}_{0,T}}\left[P^{(1)}_{\mathbbm{G}^{x,y}_{0,T}}(X) \cdot P^{(2)}_{\mathbbm{G}^{x,y}_{0,T}}(X) \cdot P^{(3)}_{\mathbbm{G}^{x,y}_{0,T}}(X)\cdot H(X)\,\mvbar\, \{X_{\psi_i}\}^{N_T}_{i=1},N_T,  \{\psi_i\}^{N_T}_{i=1}, \{\delta_i\}^{N_T}_{i=1}\right]\label{eq:pg2} \\
& = \mathbbm{E}_{\mathcal{J}} \mathbbm{E}_{\mathcal{W\,|\,J}}\mathbbm{E}_\mathcal{R}\mathbbm{E}_{\mathbbm{\overline{W}}^{x,y'}_{0,T}|R} \left[P^{(1)}_{\mathbbm{G}^{x,y}_{0,T}}(X) \cdot P^{(2)}_{\mathbbm{G}^{x,y}_{0,T}}(X) \cdot  P^{(3)}_{\mathbbm{G}^{x,y}_{0,T}}(X)\cdot H(X)\mvbar \{X_{\psi_i}\}^{N_T}_{i=1},N_T,  \{\psi_i\}^{N_T}_{i=1}, \{\delta_i\}^{N_T}_{i=1}\right]. \nonumber
\end{align}
Note that our acceptance probability $P_{\mathbbm{G}^{x,y}_{0,T}}(X)$ has been decomposed into three separate acceptance probabilities (all of which need to be accepted). This construction leads naturally to an accelerated rejection sampling strategy in which we have a sequence of acceptance probabilities and only proceed to evaluate the next conditional on acceptance of the current. $P^{(1)}_{\mathbbm{G}^{x,y}_{0,T}}(X)$ can be evaluated following the simulation of the compound Poisson process (\ref{eq:pg1}), and $P^{(2)}_{\mathbbm{G}^{x,y}_{0,T}}(X)$ can be evaluated once the trajectory of the sample path at the jump times is simulated (\ref{eq:pg2}). This leaves $P^{(3)}_{\mathbbm{G}^{x,y}_{0,T}}(X)$ which has the following form,
\begin{align}
P^{(3)}_{\mathbbm{G}^{x,y}_{0,T}}(X)
& = \prod^{N_T+1}_{i=1} e^{\Phi (\psi_i-\psi_{i-1})}\cdot\exp\left\{-\int^{\psi_i}_{\psi_{i-1}} \left[\phi(X_{s-})+\lambda(X_{s-})\right] \ud s\right\}. \label{eq:pgtilde}
\end{align}
Noting that between any two jump times with known end points that no further jumps occur and the sample path is a Brownian bridge, then each component of (\ref{eq:pgtilde}) can be considered directly using the methodology developed \secref{s:cauea}. In particular, recalling that $\phi(X_{[\psi_{i-1},\psi_i]})$ is bounded on compact sets, $\lambda(X_{[\psi_{i-1},\psi_i]})\in[0,\Lambda]$, denoting $R_i:=R_{X[\psi_{i-1},\psi_i]}\sim\mathcal{R}$ as the simulated layer (used to compute $U_i:=U_{X[\psi_{i-1},\psi_i]}\in\mathbbm{R}$ and $L_i:=L_{X[\psi_{i-1},\psi_i]}\in\mathbbm{R}$ respectively) then we can compute unbiasedly the required acceptance probability in finite computation by means of the following theorem,
\begin{theorem}[Conditioned Jump Exact Algorithm Acceptance Probability] \label{thm:acceptprob4}
Letting $\mathbbm{K}_{R(i)}$ be the law of \linebreak $\kappa(i) \sim \Poi((\Lambda\!+\! U_i\!-\!L_i)\cdot(\psi_i\!-\!\psi_{i-1}))$ and $\mathbbm{U}_{\kappa(i)}$ the distribution of $(\xi_{i,1},\ldots{},\xi_{i,\kappa(i)})\overset{\text{iid}}{\sim} \U[\psi_{i-1},\psi_i]$ we have,
\begin{align}
& {{P}}^{(3)}_{\mathbbm{G}^{x,y}_{0,T}}(X)
 = \prod^{N_T}_{i=1} \left(e^{-(L_i-\Phi)T} \cdot\mathbbm{E}_{\mathbbm{K}_{R(i)}}\left[\mathbbm{E}_{\mathbbm{U}{\kappa(i)}}\left[\prod^{\kappa(i)}_{j=1}\left(\dfrac{\Lambda+U_i\!-\!\phi(X_{\xi_j})\!-\!\lambda(X_{\xi_j})}{\Lambda+U_i\!-\!L_i}\right)\,\mvbar\,X\right]\mvbar\,X\right]\right). \nonumber
\end{align}
\end{theorem}
\noindent Simulating a finite dimensional proposal sample path as suggested above leads to the \textit{Conditioned Unbounded Jump Exact Algorithm (CUJEA)} (which for conciseness is omitted and can be found in \cite[Algorithm 5.4.1]{PhD:P13}). However, incorporating the ideas of the CAUEA of \secref{s:cauea} (\algref{alg:cauea}), leads directly to the \textit{Conditioned Adaptive Unbounded Jump Exact Algorithm (CAUJEA)} presented in \algref{alg:caujea}, outputting skeletons of the form in (\ref{eq:caujeaskel}). In \figref{fig:ill:caujea}  we present an illustrative example of an accepted CAUJEA sample path skeleton.
\begin{align}
\mathcal{S}_\text{CAUJEA}\left(X\right) &:= \bigcup^{N_T+1}_{i=1}\left\{\left(\xi_{i,j},X_{\xi_{i,j}}\right)^{\kappa(i)+1}_{j=0},\left(R^{[\xi_{i,j-1},\xi_{i,j}]}_{X[\psi_{i-1},\psi_i]}\right)^{\kappa(i)+1}_{j=1}\right\}. \label{eq:caujeaskel}
\end{align}
\begin{algorithm}[h]
	\caption{Conditioned Adaptive Unbounded Jump Exact Algorithm (CAUJEA).} \label{alg:caujea}
	\begin{enumerate}
	    \item Simulate compound Poisson process $J_{[0,T]}\sim\mathcal{J}$ as per \cite[\textsection 2.9.3]{PhD:P13}. \label{alg:cujea:init} \label{alg:caujea:init}
   	\item With probability $(1-P^{(1)}_{\mathbbm{G}^{x,y}_{0,T}}(X))$ reject path and return to \stepref{alg:cujea:init}. \label{alg:cujea:prereject1}
	\item Simulate $X'_{\psi_1},\ldots{},X'_{\psi_{N_T}}\sim\mathbbm{\overline{W}}^{x,y'}_{0,T}$ as per \cite[\textsection 2.8]{PhD:P13}. \label{alg:cujea:jumpskel}
	\item With probability $(1-P^{(2)}_{\mathbbm{G}^{x,y}_{0,T}}(X))$ reject path and return to \stepref{alg:cujea:init}. \label{alg:cujea:prereject2}

	\item For $i$ in $1$ to $(N_T+1)$, \label{alg:caujea:bisection}
	\begin{enumerate}
	\item Simulate initial layer information $R_i\sim\mathcal{R}$ as per \cite[\textsection 8.1]{B:PJR15}, setting $\Pi := \left\{\Xi\right\} := \{\{[\psi_{i-1},\psi_i], X_{\psi_{i-1}}, X_{\psi_i}, R_i\}\}$ and $\kappa_i=0$. \label{alg:caujea:layer}
    \item With probability $(1-\exp\{-(L_{X[\psi_{i-1},\psi_i]}-\Phi)\cdot (\psi_i-\psi_{i-1})\})$ reject path and return to \stepref{alg:caujea:init}.
	\item Set $\Xi=\Pi(1)$. \label{alg:caujea:loopback}
	\item Simulate $\tau\sim \text{Exp}\left(2[\Lambda+U^{\Xi}_X-L^{\Xi}_X]\right)$. If $\tau > d(\Xi)$ then set $\Pi:=\Pi\setminus\Xi$ else,
	\begin{enumerate}
	\item Set $\kappa_i=\kappa_i+1$ and with probability $1/2$ set $\xi'_{\kappa_i}=m_{\Xi}-\tau$ else $\xi'_{\kappa_i}=m_{\Xi}+\tau$.
	\item Simulate $X_{\xi'_{\kappa_i}}\sim\left.\mathbbm{\overline{W}}^{x(\Xi), y(\Xi)}_{\bar{s}(\Xi),\bar{t}(\Xi)}\,\mvbar\,R^{\Xi}_i\right.$ as per \cite[\textsection 8.2]{B:PJR15}. \label{alg:caujea:intermediate}
	\item With prob. $(1\!-\![\Lambda\!+\!U^{\Xi}_{X}\!-\!\phi(X_{\xi'_{\kappa_i}})\!-\!\lambda(X_{\xi'_{\kappa_i}})]/[\Lambda\!+\!U_{X}\!-\!L^{\Xi}_{X}])$ reject path and return to \stepref{alg:caujea:init}.
	\item Simulate new layer information $R_i^{[\bar{s}(\Xi),\xi'_{\kappa_i}]}$ and $R_i^{[\xi'_{\kappa_i},\bar{t}(\Xi)]}$ conditional on $R^{\Xi}_i$ as per \cite[\textsection 8.3 \& \textsection 8.4]{B:PJR15}. \label{alg:caujea:newlayer}
	\item With probability $\Big(1-\exp\Big\{-\Big[L^{[\bar{s}(\Xi),\xi'_{\kappa_i}]}_{X[\psi_{i-1},\psi_i]}+L^{[\xi'_{\kappa_i},\bar{t}(\Xi)]}_{X[\psi_{i-1},\psi_i]}-2L^{\Xi}_{X[\psi_{i-1},\psi_i]}\Big]\cdot[d_{\Xi}-\tau]\Big\}\Big)$ reject path and return to \stepref{alg:caujea:init}.
	\item Set  $\Pi:=\Pi\bigcup\Big\{[s_{\Xi},m_{\Xi}-\tau], X^{\Xi}_{\bar{s}}, X_{\xi'_{\kappa_i}}, R^{[\bar{s}(\Xi),\xi'_{\kappa_i}]}_i\Big\}\bigcup\Big\{[m_{\Xi}+\tau,t_{\Xi}],X_{\xi'_{\kappa_i}}, X^\Xi_{\bar{t}}, R^{[\xi'_{\kappa_i},\bar{t}(\Xi)]}_i\Big\}\setminus\Xi$.
	\end{enumerate}
	\item If $\big|\Pi\big|\neq 0$ return to \stepref{alg:caujea:loopback}. 
	\item Define skeletal points $\xi_{i,1},\ldots{},\xi_{i,\kappa_i}$ as the order statistics of the set $\{\xi'_{i,1},\ldots{},\xi'_{i,\kappa_i}\}$.
	\end{enumerate}
	\item Accept sample path skeleton.\\
	\vspace{-0.25cm}\hrule\vspace{0.15cm}
	\item \textit{* Simulate $\rem{X}\sim\left(\otimes^{N_T+1}_{i=1}\left(\otimes^{\kappa_i+1}_{j=1}\mathbbm{W}^{X_{\xi_{i,j-1}},X_{\xi_{i,j}}}_{\xi_{i,j-1},\xi_{i,j}}\,\mvbar\,R^{[\xi_{i,j-1},\xi_{i,j}]}_i\right)\right)$}. 
	\vspace{-0.35cm}
	\end{enumerate}
\end{algorithm}

\FloatBarrier
\section*{ACKNOWLEDGMENTS}
MP would like to thank Fl{\'a}vio Gon{\c{c}}alves, Adam Johansen and Gareth Roberts for stimulating discussion on this paper. This work was supported by the EPSRC [grant numbers EP/P50516X/1 and EP/K014463/1].

\bibliographystyle{arxiv}
\bibliography{BibtexReferences}

\section*{AUTHOR BIOGRAPHY}

\noindent {\bf MURRAY POLLOCK} is a Postdoctoral Research Fellow in Statistics based at the University of Warwick working on the EPSRC programme grant ``Intractable Likelihood: New Challenges from Modern Applications (i-like),'' held jointly along with Bristol, Lancaster, and Oxford universities. His research interests lie in Monte Carlo methodology (particularly MCMC and SMC). His email address is \email{m.pollock@warwick.ac.uk}.

\end{document}